\documentstyle[12pt]{article}
\newcommand\beq{\begin{equation}}
\newcommand\eeq{\end{equation}}
\newcommand\bea{\begin{eqnarray}}
\newcommand\eea{\end{eqnarray}}
\newcommand\pa{\partial}

\newcommand\ti{\tilde}

\newcommand{\cl}{\centerline}

\begin{document}
\begin{titlepage}
\setlength{\textwidth}{5.0in}
\setlength{\textheight}{7.5in}
\setlength{\parskip}{0.0in}
\setlength{\baselineskip}{18.2pt}
\hfill
{\tt HD-THEP-02-9, SOGANG-HEP 292/02}
\begin{center}
{\large{\bf Symplectic quantization of self-dual master Lagrangian}}\par
\vskip 0.3cm
\begin{center}
{Soon-Tae Hong$^{1}$, Yong-Wan Kim$^{1}$, Young-Jai Park$^{1}$ and
Klaus D. Rothe$^{2}$}\par
\end{center}
\vskip 0.3cm
\begin{center}
{$^{1}$Department of Physics and Basic Science Research Institute,}\par
{Sogang University, C.P.O. Box 1142, Seoul 100-611, Korea}\par
{$^{2}$Institut f\"ur Theoretische Physik,}\par
{Universit\"at Heidelberg, Philosophenweg 16, D-69120 Heidelberg, Germany}\par
\end{center}
\cl{\today}
\vskip 0.2cm
\begin{center}
{\bf ABSTRACT}
\end{center}
\begin{quotation}
We consider the master Lagrangian of Deser and Jackiw, interpolating between the self-dual and the
Maxwell-Chern-Simons Lagrangian, and quantize it following the symplectic approach, as well as the traditional
Dirac scheme. We demonstrate the equivalence of these procedures in the subspace of the second-class constraints.
We then proceed to embed this mixed first- and second-class system into an extended first-class system within the
framework of both approaches, and construct the corresponding generator for this extended gauge symmetry in both
formulations. \vskip 0.2cm \noindent
PACS: 11.10.Ef, 11.10.Kk, 11.15.-q\\
\noindent
Keywords: symplectic quantization, second-class, self-dual, local symmetries,
Lagrangian approach\\
\noindent
\end{quotation}
\end{center}
\end{titlepage}

\newpage

\section{Introduction}

The traditional Dirac quantization method (DQM)~\cite{dirac64} has been widely used in order to quantize
Hamiltonian systems involving first- and second-class constraints. The resulting Dirac brackets defined on the
subspace of the constraints may however be field-dependent and nonlocal, and could thus pose serious ordering
problems for the quantization of the theory.  On the other hand, the Becci-Rouet-Stora-Tyutin
(BRST)~\cite{becci76,kugo79} procedure of first turning the second-class constraints into first-class ones along
the lines originally established by Batalin, Fradkin, and Vilkovisky ~\cite{fradkin75,henneaux85}, and then
reformulated in a more tractable and elegant version by Batalin, Fradkin, and Tyutin (BFT)~\cite{batalin87}, does
not suffer from these difficulties, as it relies on a simple Poisson bracket structure. As a result, the embedding
of second-class systems into first-class ones (gauge theories) has received much attention in the past years and
the DQM improved in this way, has been applied to a number of
models~\cite{idqm,KimRothe,nonPro,gafn,fujiwara90,kim92,banerjee94,kim95, KimParkRothe,banerjee95ap,kkky,phyrep}
in order to obtain the corresponding Wess-Zumino (WZ) actions~\cite{faddeev86,wess71}.

The traditional Dirac approach~\cite{dirac64} has been criticized for introducing ``superfluous'' (primary)
constraints. As a result an alternative approach based on the symplectic structure of phase space has been
proposed in ref. \cite{jackiw85}. The advantage of such an approach in the case of first-order Lagrangians such as
Chern-Simons theories has in particular been emphasized by Faddeev and Jackiw~\cite{jackiw85}. This symplectic
scheme has been worked out in considerable detail in a series of papers \cite{wozneto}, and has been applied to a
number of models~\cite{wozneto,kimjkps1}. It has further been extended recently to implement the improved DQM
embedding program in the context of the symplectic formalism~\cite{kimjkps2,neto0109,HKPR}.

In this paper, we wish to illustrate the above quantization schemes in the case of the self-dual master Lagrangian
of Deser and Jackiw~\cite{deser}.  The material is organized as follows. In section 2, we briefly discuss the
self-dual master model within the framework of the standard and the improved DQMs. In section 3, we apply the
gauge non-invariant symplectic formalism~\cite{jackiw85,wozneto} to this model. In section 4, we then show how the
improved DQM program for gauging all degrees of freedom in this master Lagrangian is realized in the framework of
the symplectic formalism. We also briefly discuss the one-to-one correspondance with the traditional Dirac and
improved Dirac approach in the respective cases. Our conclusion is given in section 5.

\section{Dirac quantization method}
\setcounter{equation}{0}
\renewcommand{\theequation}{\arabic{section}.\arabic{equation}}
\begin{center}
{\bf Standard Dirac quantization method}
\end{center}

In this section, we consider the massive self-dual model
Lagrangian~\cite{deser}
\begin{equation}
\label{action} {\cal L}_{0}= \frac{m}{2}f_{\mu} f^{\mu}
-\frac{1}{2}\epsilon_{\mu\nu\lambda}f^{\mu}\pa^{\nu}A^{\lambda}
-\frac{1}{2}\epsilon_{\mu\nu\lambda}A^{\mu}\pa^{\nu}f^{\lambda}
+\frac{1}{2}\epsilon_{\mu\nu\lambda}A^{\mu}\pa^{\nu}A^{\lambda}.
\end{equation}
The canonical momenta conjugate to the fields $f^{\mu}$ and
$A^{\mu}$ are given by
\bea
\label{canmom}
\pi_{0}^{f}&=& 0,~~~\pi_{i}^{f}= -\frac{1}{2}\epsilon_{ij}A^{j},\nonumber\\
\pi_{0}^{A}&=& 0,~~~\pi_{i}^{A}=-\frac{1}{2}\epsilon_{ij}f^{j} +\frac{1}{2}\epsilon_{ij}A^{j} \eea with the
Poisson algebra $\{f^{\mu}(x), \pi_{\nu}^{f}(y) \} =\{A^{\mu}(x), \pi_{\nu}^{A}(y) \}=\delta ^{\mu}_{\nu}
\delta(x-y)$. The canonical Hamiltonian then reads
\begin{equation}
\label{canH} H_c= \int d^2x~ \left[-\frac{m}{2}f_{\mu}f^{\mu}+\epsilon_{ij}f^{0}\pa^{i}A^{j}
+\epsilon_{ij}A^{0}(\pa^{i}f^{j}-\pa^{i}A^{j})\right].
\end{equation}
The primary constraints following from the definition of the canonical momenta,
are
\bea \label{pricon}
\phi^f_0 &\equiv& \pi_{0}^{f} \approx 0,\nonumber \\
\phi^f_i &\equiv& \pi_{i}^{f} + \frac{1}{2}\epsilon_{ij}A^{j} \approx 0,\nonumber\\
\phi^A_0 &\equiv& \pi_{0}^{A} \approx 0, \nonumber \\
\phi^A_i &\equiv& \pi_{i}^{A}+\frac{1}{2}\epsilon_{ij}f^{j} -\frac{1}{2}\epsilon_{ij}A^{j} \approx 0 \eea with the
corresponding primary Hamiltonian $H_p$ \beq \label{hprimary} H_p=H_c+\int d^2x~ \sum_{\mu=0}^2(v^\mu_f\phi^f_\mu
+v^\mu_A \phi^A_\mu). \eeq Persistence in time of the primary constraints leads to the secondary constraints
\bea\label{secondaryconstr}
\varphi^f &\equiv& m f^0-\epsilon_{ij}\partial^iA^j \approx 0, \nonumber \\
\varphi^A &\equiv& -\epsilon_{ij}(\partial^if^j-\partial^iA^j)\approx 0. \eea The constraints $\phi^f_i$ and
$\phi^A_i$ fix the corresponding Lagrange multipliers to be $v^i_f= \partial^if^0+m\epsilon^{ij}f_j$ and $v^i_A=
\partial^i A^0+ m\epsilon^{ij}f_j$, respectively. The Lagrange multiplier $v^0_f$ is determined by the time evolution
of the constraint $\varphi^f$ with the primary Hamiltonian to be $v^0_f=\partial_i f^i$, while the multiplier
$v^0_A$ remains undetermined. The fact of having an undetermined Lagrange multiplier $v^0_A$ reflects the
existence of a gauge symmetry related to the fields $A^\mu$. Indeed, with the redefinition \cite{BanerjeeRothe} of
the constraint $\varphi^A \to \omega^A = \varphi^A + \partial^i{\phi^A_i}$, we see that $\omega^A$ is first class
and is the generator of the  gauge transformation, $A^i \rightarrow A^i + \partial^i \lambda$.

We could now follow the BFT procedure in order to turn all constraints into first-class ones. This is left for the
appendix. Here we are primarily interested in establishing the connection between the BFT embedding and the
symplectic embedding procedures~\cite{HKPR}. As it turns out (see section 3 and 4), this connection is given in
the subspace where the constraints  $\phi^f_i=0$ and $\phi^A_i=0$ are implemented strongly. In this subspace we
are then left with two first-class constraints, $\phi^A_0\approx 0$, $\varphi^A\approx 0$, and two second-class
ones, $\phi^f_0\approx 0$, $\varphi^f\approx 0$. We construct the corresponding Dirac brackets in terms of the
inverse of the matrix $\Delta$ defined in terms of the Poisson brackets of $\{ \phi^f_i, \phi^A_j\}$: \beq
\Delta(x,y) = \left(
\begin{array}{cc}
0&\epsilon\\
\epsilon&-\epsilon
\end{array}
\right)\delta^2(x-y),
\eeq
where
\beq
\label{epsilon}
\epsilon = \left(
\begin{array}{cc}
0&1\\
-1&0
\end{array}
\right).
\eeq
For the  corresponding non-vanishing Dirac brackets, computed in the standard way, we have:
\bea\label{Diracbrackets}
\{A^0, \pi^A_0 \}_D&=& \delta^2(x-y), ~~~~~~~~~~~~\{f^0, \pi^f_0 \}_D= \delta^2(x-y),\nonumber \\
\{f^i, f^j \}_D&=-& \epsilon^{ij}\delta^2(x-y),
 ~~~~~~~~\{f^i, A^j \}_D= -\epsilon^{ij}\delta^2(x-y),\nonumber\\
\{A^i, \pi^A_j \}_D&=& \frac{1}{2}\delta^i_j\delta^2(x-y),
~~~~~~~\{f^i, \pi^f_j \}_D= \frac{1}{2}\delta^i_j\delta^2(x-y),\nonumber \\
\{f^i, \pi^A_j \}_D&=&\frac{1}{2}\delta^i_j\delta^2(x-y), \eea where we have used the convention:
$\epsilon_{12}=\epsilon^{12}=1$ and $\epsilon_{ik}\epsilon^{kj}=-\delta^j_i$.

 \vskip 0.5cm
 \newpage
\begin{center}
{\bf Improved Dirac Quantization Method}
\end{center}

Following the Improved
DQM~\cite{idqm,KimRothe,nonPro,gafn,fujiwara90,kim92,banerjee94,kim95,KimParkRothe,banerjee95ap,kkky}, we now
proceed to embed the model into a gauge theory with respect to the above Dirac brackets, by extending phase space
to include a pair of (canonically conjugate) auxiliary fields $\Phi^i$, satisfying the Poisson brackets
\begin{equation}
\label{intro-aux} \{\Phi^i(x), \Phi^j(y)\}=\epsilon^{ij}\delta^2(x-y).
\end{equation}
Denote the second-class constraints $(\phi^f_0,\varphi^f)$ by $\{\Omega^f_i\}, i=1,2$. The first-class constraints
$\tilde{\Omega}^f_i$ are now constructed as a power series in the auxiliary fields, as follows:
\begin{equation}
\label{ansatz}
\tilde{\Omega}^f_i = {\Omega}^f_i + \sum^{\infty}_{n=1} \Omega^{(n)}_i,
\end{equation}
where $\Omega^{(n)}_i , (n=1,\cdots,\infty)$ are homogeneous polynomials in the auxiliary fields $\Phi^i$ of
degree $n$, to be determined by the requirement that the constraints $\tilde{\Omega}^f_i$ be strongly involutive:
\beq\label{involution} \{\tilde{\Omega}^f_i(x),\tilde{\Omega}^f_j(y)\}_D = 0. \eeq Making the ansatz,
\beq\label{ansatzlinear} \Omega^{(1)}_i(x) = \int d^2y X_{ij}(x,y)\Phi^j(y) \eeq and substituting this ansatz into
(\ref{ansatz}), the requirement (\ref{involution}) leads to the simple solution
$X_{ij}(x,y)=\sqrt{m}\delta_{ij}\delta^2(x-y)$. There are no higher order contributions to (\ref{ansatz}). We thus
obtain for the first-class constraints \bea \tilde{\phi}^f_{0}&=&\pi_{0}^{f}+m\theta,
\nonumber\\
\tilde\varphi^f&=& mf^{0}-\epsilon_{ij} \pa^{i}A^{j}+\pi_{\theta}\,, \label{consttil} \eea satisfying the
first-class algebra $\{\tilde{\phi}^f_{0}, \tilde\varphi^f \}=0$, where we have replaced ($\Phi^1,\Phi^2$) by
($\sqrt{m}\theta$, $\pi_\theta/\sqrt{m})$, for convenience.

Applying this procedure to the original field variables, we similarly obtain for the corresponding first-class
fields \bea
\tilde{f}^{0}&=& f^{0}+\frac{1}{m}\pi_\theta, \nonumber \\
\tilde{f}^{i}&=& f^{i}+\pa^{i}\theta, \nonumber\\
\tilde{\pi}^f_0&=&\pi^f_0+ m\theta,  \eea satisfying $\{ F(\tilde{f}^\mu, \tilde{\pi}^f_0), \tilde{\Omega}^f_i
\}=0$.

Since an arbitrary functional of the first-class fields is also first-class, we can obtain the first-class
Hamiltonian ${\tilde H}_c$ by simply replacing the original fields by the respective tilde-fields
\cite{KimParkRothe,gr}: \bea \tilde{H}_{c} &=&\int d^2x
\left[-\frac{m}{2} (f^{0}+\frac{1}{m}\pi_{\theta})^{2} +\frac{m}{2}(f^{i}+\pa^{i}\theta)^{2} \right. \nonumber \\
&& \left. +\epsilon_{ij}(f^{0}+\frac{1}{m}\pi_\theta)\pa^{i}A^{j} +
\epsilon_{ij}A^{0}(\pa^{i}f^{j}-\pa^{i}A^{j})\right] \label{Hfirstclass} \eea
along with the remaining first-class
constraints now written in the extended phase space as \bea \label{firA}
\tilde{\phi}^A_0 &=& \tilde{\pi}^A_0 = \pi^A_0, \nonumber \\
\tilde{\varphi}^A &=& - \epsilon_{ij}\pa^i\tilde{f}^j+\epsilon_{ij}\pa^i\tilde{A}^j =
-\epsilon_{ij}\pa^if^j+\epsilon_{ij}\pa^iA^j.
\eea

Note that we have taken here $\tilde{A}^\mu = A^\mu$. Indeed, $A^\mu$ and $\pi^A_0$ remain unchanged by the
embedding procedure, which only involved the $f^\mu$-fields. $A^\mu$ thus continues to transform as usual under
gauge transformations, and is not first class. One may thus question our simple substitution procedure for
arriving at the first-class Hamiltonian. It is thus instructive to construct $\tilde{H}_c$ following the usual BFT
construction in order to obtain the involutive Hamiltonian directly~\cite{batalin87}. The procedure assumes that
the involutive Hamiltonian can be written as the infinite series
\begin{equation}
\tilde{H}=H_c+\sum^{\infty}_{n=1}H^{(n)}, ~~H^{(n)}\sim (\Phi^i)^n,
\end{equation}
satisfying the initial condition $\tilde{H}(f^\mu,A^\mu,\pi^f_0,\pi^A_0;\Phi^i=0)=H_c$. The general
solution~\cite{batalin87} for the involutive Hamiltonian $\tilde{H}$ is then given by
\begin{equation}
\label{genInvSol}
H^{(n)}=-\frac{1}{n}\int d^2x d^2y~\Phi^i(x)\epsilon_{ij}X^{jk}(x,y)G^{(n-1)}_k(y),
\end{equation}
where the generating functionals $G^{(n)}$ are:
\begin{eqnarray}
&& G^{(0)}_i=\{\Omega^{(0)}_i, H_c\}, \nonumber\\
&& G^{(n)}_i=\{\Omega^{(0)}_i, H^{(n)}\}_{\cal O}+\{\Omega^{(1)}_i, H^{(n-1)}\}_{\cal O}.
\end{eqnarray}
Here the symbol ${\cal O}$ denotes that the Poisson brackets are calculated among the original variables.

Explicit calculations for our model yield
\begin{equation}
G^{(0)}_1 = mf^0-\epsilon_{ij}\partial^iA^j,~~G^{(0)}_2=m\partial_i f^i,
\end{equation}
which are substituted in (\ref{genInvSol}) to obtain $H^{(1)}$:
\begin{equation}
H^{(1)}=\int d^2x~ \left[ m\theta\partial_if^i-\frac{1}{m}\pi_\theta(mf^0-\epsilon_{ij}\partial^iA^j)\right].
\end{equation}
The generating functionals for the next generation are:
\begin{eqnarray}
G^{(1)}_1 = \pi_\theta,~~G^{(1)}_2=m\partial_i \partial^i\theta,
\end{eqnarray}
and yield
\begin{equation}
H^{(2)}=\int d^2x~\left[-\frac{1}{2m}\pi^2_\theta -\frac{m}{2}\partial_i\theta\partial^i\theta \right].
\end{equation}
There are no further iterative higher order Hamiltonians, and thus total Hamiltonian can be written as
\begin{eqnarray}
\label{invH}
\tilde{H}&=&H_c+H^{(1)}+H^{(2)} \nonumber\\
         &=&\int d^2x \left[ -\frac{m}{2}f_\mu f^\mu+f^0\epsilon_{ij}\partial^iA^j
               +A^0\epsilon_{ij}(\partial^if^j-\partial^iA^j)\right. \nonumber \\
         && + \left. m\theta\partial_i f^i-f^0\pi_\theta+\frac{1}{m}\pi_\theta\epsilon_{ij}\partial^iA^j
             -\frac{1}{2m}\pi^2_\theta-\frac{m}{2}\partial_i\theta\partial^i\theta \right],
\end{eqnarray}
which is the same as the first-class Hamiltonian (\ref{Hfirstclass}) up to a total derivative. This confirms the
equivalence of the $\tilde{H}_c$ (\ref{Hfirstclass}) of the involutive Hamiltonian $\tilde{H}$ (\ref{invH}).


Now, let us streamline the notation by defining \bea
(\tilde{\Omega}^f_\alpha) &=& (\tilde\phi^f_0,\tilde\varphi^f)\,,\nonumber\\
(\tilde{\Omega}^A_\alpha) &=& (\tilde\phi^A_0,\tilde\varphi^A). \eea With respect to the Dirac brackets defined
previously we then have the relations of strong involution
\begin{eqnarray}
&& \{\tilde{\Omega}^f_\alpha,\tilde{\Omega}^A_\beta\}_D =0, \nonumber \\
&& \{\tilde{\Omega}^f_\alpha,\tilde{H}_c\}_D =0, \nonumber \\
&& \{\tilde{\phi}^A_0,\tilde{H}_c\}_D =\tilde{\varphi}^A, \nonumber \\
&& \{\tilde{\varphi}^A,\tilde{H}_c\}_D =0,
\end{eqnarray}

On the other hand, with the first-class Hamiltonian (\ref{Hfirstclass}), one does not generate naturally the
first-class Gauss law constraints from the time evolution of the primary constraints $\tilde{\phi}^f_0 \approx
0\,,~\tilde{\phi}^A_0 \approx 0$. For this to be the case we introduce an additional term proportional to the
first class constraints $\tilde\phi^f_0$ into the Hamiltonian density $\tilde{\cal H}_{c}$, leading us to consider
the equivalent first-class Hamiltonian
\begin{equation}
\tilde{\cal H}_{c}^{\prime}=\tilde{\cal H}_{c}+\frac{1}{m}\pi_{\theta}\tilde{\phi}^f_0\,. \label{hctp}
\end{equation}
We then obtain the Dirac brackets in the desired form:
\begin{eqnarray}
&& \{\tilde{\phi}^f_0(x),\tilde{H}_{c}^{\prime}\}_D =\tilde{\varphi}^f(x)\,,\quad
\{\tilde{\varphi}^f(x),\tilde{H}_c^\prime\} = 0,
\nonumber \\
&& \{\tilde{\phi}^A_0(x),\tilde{H}_{c}^{\prime}\}_D =\tilde{\varphi}^A(x)\,,\quad
\{\tilde{\varphi}^A(x),\tilde{H}_c^\prime\} = 0. \label{ga}
\end{eqnarray}

We streamline further the notation by collecting all the first-class constraints into a single vector:
\beq\label{firstclassconstr} \tilde\Omega_A = (\tilde\Omega^f_\alpha,\tilde\Omega^A_\alpha). \eeq Note here that
the subscript $A$ is the index running 1 to 4, while the superscript $A$ in $\tilde{\Omega}^A_\alpha$ denotes
these constraints are related to the field $A^\mu$ in the model.

We now seek the equivalent Lagrangian corresponding to the first-class Hamiltonian $\tilde{\cal H}_{c}^{\prime}$
in (\ref{hctp}).  To this end we consider the  partition function in the phase space as given by the
Faddeev-Senjanovic prescription \cite{faddev}, \bea\label{partitionfunction} && Z=N\int {\cal D}f^{\mu}{\cal
D}A^{\mu}{\cal D}\theta {\cal D}\pi_{\theta} {\cal D} \pi_{0}^{f}{\cal D}\pi_{0}^{A} \prod_{A,B}
\delta(\tilde{\Omega}_A)\delta(\Gamma_B) \det|\{\tilde {\Omega}_A,\Gamma_B\}|
e^{i\int {\rm d}^3x {\cal L}},\nonumber\\
&& {\cal L}=\pi_{0}^{f}\dot{f}^{0}-\frac{1}{2}\epsilon_{ij}A^j\dot{f}^i +\pi_{0}^{A}\dot{A}^{0}
+(-\frac{1}{2}\epsilon_{ij}f^j+\frac{1}{2}\epsilon_{ij} A^j)\dot{A}^i +\pi_{\theta}\dot{\theta} -\tilde{\cal
H}_{c}^{\prime},\nonumber
\\
\eea
where the gauge fixing conditions $\Gamma_B$ are chosen so that the determinant occurring in the functional
measure is nonvanishing.

Exponentiating the delta function $\delta (\tilde{\varphi}^f)$ as $\delta (\tilde{\varphi}^f)=\int {\cal D}\xi
e^{i\int {\rm d}^3x~\xi \tilde{\varphi}_f}$, making a transformation $f^{0}\rightarrow f^{0}+\xi$ and performing
the integration over $\pi_{0}^{f}$, $\pi_{\theta}$, $\pi^A_0$ and $\xi$, the partition function
(\ref{partitionfunction}) reduces to \beq \label{fca} Z = N \int {\cal D}f^{\mu} {\cal D}A^{\mu} {\cal D}\theta
\prod_{A,B}\delta(\Gamma_B)\det|\{\tilde{\Omega}_A,\Gamma_B\}| e^{i\int {\rm d}^3 {\cal L}}, \eeq where
\bea\label{L-Stuckelberg} {\cal L}&=&\frac{m}{2}(f_\mu+\pa_\mu\theta) (f^\mu+\pa^\mu\theta)
-\frac{1}{2}\epsilon_{\mu\nu\lambda}(f^\mu+\pa^\mu\theta)\pa^\nu A^\lambda
\nonumber\\
&-&\frac{1}{2}\epsilon_{\mu\nu\lambda}A^\mu\pa^\nu(f^\lambda+\pa^\lambda\theta)
+\frac{1}{2}\epsilon_{\mu\nu\lambda}A^\mu\pa^\nu A^\lambda \label{zlag}
\end{eqnarray}
is the manifestly gauge invariant St\"uckelberg Lagrangian with  the St\"uckelberg scalar $\theta$.

Next, following Dirac's conjecture~\cite{dirac64}, we construct the generator $G$ of gauge transformations for the
embedded self-dual master model in the standard way,
\begin{equation}
\label{generator} G=\int{\rm d}^{3}x~\sum_\alpha\left[\epsilon_\alpha^f \tilde{\Omega}_\alpha^f +
\epsilon_\alpha^A \tilde{\Omega}_\alpha^A\right],
\end{equation}
where $\epsilon^f_\alpha, \epsilon^A_\alpha$ are in general functions of phase space variables and $\tilde{\Omega}^f_\alpha, \tilde{\Omega}^A_\alpha$ are the first-class
constraints in Eq. (\ref{firstclassconstr}). The infinitesimal gauge transformation for a function $F$ of phase space variables is then given by
the relation of $\delta F= \{ F, G \}_D$, and leads to
\begin{eqnarray}
\delta f^{0}&=& \epsilon^f_1,~~~ \delta f^{i}=-\pa^{i} \epsilon^f_2, \nonumber\\
\delta A^{0}&=& \epsilon^A_1,~~~ \delta A^{i}=-\pa^{i} \epsilon^A_2, \nonumber\\
\delta \theta &=& \epsilon^f_2. \label{gtrm}
\end{eqnarray}

The above gauge transformation involving four gauge parameters is a symmetry of the Hamiltonian, but not of the Lagrangian.
The generator $G$ of the most general local symmetry transformation of a
Lagrangian must satisfy the master equation \cite{brr}
\begin{equation}\label{mastereq}
\frac{\partial G}{\partial t} + \{G, H_T \} =0,
\end{equation}
which, together with (\ref{generator}), implies the following well-known restrictions on the gauge parameters, and
on the Lagrange multipliers in the primary Hamiltonian:
\begin{eqnarray}
\label{restrictions}
 \delta v^{\beta} &=& \frac{d\epsilon^{\beta}}{dt}
-\epsilon^A(V^{\beta}_A+v^{\alpha}C^{\beta}_{\alpha A}), \nonumber \\
 0&=& \frac{d\epsilon^{b}}{dt}-\epsilon^A(V^{b}_A+v^{\alpha}C^{b}_{\alpha A}).
\end{eqnarray}
Here the superscript $\alpha,\beta$ ($a,b$) denote the primary (secondary) constraints, and  $V^A_B$,
$C^A_{BC}$ are the structure functions of the constrained Hamiltonian dynamics defined by
$\{H_c, \tilde\Omega_A \}_D=V^B_A\tilde\Omega_B$, $\{\tilde\Omega_A,
\tilde\Omega_B\}_D=C^C_{AB}\tilde\Omega_C$, respectively. From (\ref{restrictions})
we obtain $\epsilon^f_1=-d\epsilon^f_2/dt$, and
$\epsilon^A_1=-d\epsilon^A_2/dt$. Thus the gauge transformations (\ref{gtrm}) reduce
\begin{equation}
\label{idqm-transformation} \delta f^\mu = -\partial^\mu \epsilon^f_2, ~~\delta A^\mu = -\partial^\mu\epsilon^A_2,
~~\delta\theta=\epsilon^f_2,
\end{equation}
which evidently leaves the St\"uckelberg Lagrangian  (\ref{L-Stuckelberg}) invariant.

\section{Constraint structure of master Lagrangian in symplectic approach}
\renewcommand{\theequation}{\arabic{section}.\arabic{equation}}
\setcounter{equation}{0}
\renewcommand{\theequation}{\arabic{section}.\arabic{equation}}
In this and the following sections we show that the results obtained in section 2 are in full agreement with those
obtained in the symplectic approach. We begin by considering the symplectic analogue of the conventional Dirac
approach.

The Master Lagrangian (\ref{action}) is of the form \beq \label{Lsymplectic} L=\int d^2x~ a(x)_\alpha\dot
\xi_\alpha(x) - V[\xi], \eeq where \beq\label{xi0} (\xi_\alpha) = (f^1,f^2,A^1,A^2,f^0,A^0), \eeq \beq (a_\alpha)
= \left(-\frac{1}{2}A^2,\frac{1}{2}A^1,-\frac{1}{2}(f^2 - A^2), \frac{1}{2}(f^1 - A^1),0,0\right), \eeq and \beq
V=\int d^2x~\left[-{m\over 2}f_\mu f^\mu +f^0\epsilon_{ij}\partial^i A^j + A^0(\epsilon_{ij}\partial^i f^j -
\epsilon_{ij}\partial^i A^j )\right]. \label{pot} \eeq The Euler-Lagrange equations then read \beq
\label{EulerLagrange} \int d^2y~ F^{(0)}_{\alpha\beta}(x,y) \dot\xi_\beta(y) = K^{(0)}_\alpha(x)\,, \eeq where
\beq \label{Kvector} (K^{(0)}_\alpha)= \frac{\delta V}{\delta \xi_\alpha(x)} = \left(
\begin{array}{c}
\partial^2 A^0 + m f^1
\\
-\partial^1 A^0 + m f^2
\\
-\partial_2(f^0 - A^0)
\\
\partial_1(f^0 - A^0)
\\
\epsilon_{ij}\partial^iA^j - m f^0
\\
\epsilon_{ij}\partial^i(f^j - A^j)
\end{array}\right),
\eeq
and $F^{(0)}_{\alpha\beta}$ is the (pre)symplectic form~\cite{wozneto}
\beq
\label{Fpresymplectic}
F^{(0)}_{\alpha\beta}(x,y) = \frac{\partial a_\beta(y)}{\partial \xi_\alpha(x)} - \frac{\partial
a_\alpha(x)}{\partial\xi_\beta(y)}. \eeq
Explicitly
\beq F^{(0)}(x,y)=\left(
\begin{array}{ccc}
0 & \epsilon & 0
\\
\epsilon & -\epsilon & 0
\\
0 & 0 & 0
\end{array}\right)\delta^2(x-y),
\eeq where $\epsilon$ is the matrix (\ref{epsilon}), and $0$ is the $2\times2$ matrix. It is evident that since
$\det F^{(0)}=0$, the matrix $F^{(0)}$ is not invertible. In fact, the rank of this matrix is four, so that there
exist two-fold infinity of zero-generation (left) zero modes $u^{(0)}(\sigma;z)$, labelled by discrete indices
$\sigma=1,2$ and the continuum label $z$, with components:
\bea u^{(0)T}_{x}(1;z)&=&(0,0,0,0,-1,0)\delta^2(x-z),
\nonumber\\
u^{(0)T}_{x}(2;z)&=&(0,0,0,0,0,-1)\delta^2(x-z),
\eea
where the superscript ``{\it T}" stands for ``transpose".
Correspondingly we have a two-fold infinity of ``zero generation" constraints
\beq \varphi_\sigma(z)=\int d^2x
\sum_\alpha u^{(0)}_{\alpha,x}(\sigma,z) \frac{\delta V}{\delta\xi_\alpha(x)}=0\,.
\eeq
Explicitly
\bea
\varphi_1(z)&=&-\frac{\delta V}{\delta f^0(z)} =m f^0(z) - \epsilon_{ij}\partial^iA^j(z),
\nonumber\\
\varphi_2(z)&=&-\frac{\delta V}{\delta A^0(z)} =-\epsilon_{ij}\partial^i(f^j(z) - A^j(z)). \label{phi1phi2}
\eea
Comparing with (\ref{secondaryconstr}) we see that $\varphi_1 = \varphi^f$ and  $\varphi_2 = \varphi^A$. We must
require these constraints to be conserved in time:
\beq
\partial_0\varphi
_\sigma(z)=0\,,
\eeq
or
\beq \int d^2 x \sum_\alpha\frac{\partial\varphi_\sigma(z)}{\partial\xi_\alpha(x)}
\dot\xi_\alpha(x)= 0\,.
\eeq
These equations of motion are obtained as one of the Euler-Lagrange equations of the
extended Lagrangian\footnote{The minus sign is chosen for later convenience, when comparing with the Dirac
quantization procedure.}
\beq L' = L - \int d^2x\sum_\sigma\varphi_\sigma(z)\dot\eta_\sigma(z).
\eeq
The field
$A^0$ only occurs in the potential $V$ in the form $A^0(z)\varphi_2(z)$. Hence it can be absorbed into a new
dynamical variable  via the shift $\dot\eta_2 -A^0 \to \dot\eta_2$. Our new set of ``first-generation" dynamical
variables are then
\beq (\xi^{(1)}_{\alpha_1}) = (f^1,f^2,A^1,A^2,f^0,\eta_1,\eta_2),
\eeq
and the ``first-generation" Lagrangian takes the form
\beq L^{(1)} = \int d^2x
\sum_{\alpha_1=1}^7a^{(1)}_{\alpha_1}\dot\xi^{(1)}_{\alpha_1} - V^{(1)}[\xi], \label{firstlag}
\eeq
where
\beq
\left( a^{(1)}_{\alpha_1}(x) \right) = \left(-\frac{1}{2}A^2,\frac{1}{2}A^1,-\frac{1}{2}(f^2 - A^2),
\frac{1}{2}(f^1 - A^1),0,-\varphi_1(x),-\varphi_2(x)\right),
\eeq
and
\beq V^{(1)}=\int d^2x~\left[-{m\over
2}f_\mu f^\mu + f^0\epsilon_{ij}\partial^i A^j\right]. \label{pot1}
\eeq
The equations of motion now take the form
\beq\label{firstlevel-eq} \int d^2y~ F^{(1)}_{\alpha_1,\beta_1}(x,y)\dot\xi^{(1)}_{\beta_1}(y) = \frac{\delta
V^{(1)}}{\delta\xi^{(1)}_{\alpha_1}(x)},
\eeq
where the ``first-generation" symplectic form
$F^{(1)}_{\alpha_1,\beta_1}$ is given by
\beq\label{F1-symplectic} F^{(1)}_{\alpha_1\beta_1}(x,y) = \frac{\partial
a_{\beta_1}^{(1)}(y)}{\partial \xi_{\alpha_1}^{(1)}(x)} - \frac{\partial
a_{\alpha_1}^{(1)}(x)}{\partial\xi_{\beta_1}^{(1)}(y)}\,,
\eeq
or explicitly
\beq F^{(1)}(x,y)=\left(
\begin{array}{ccccccc}
0&0&0&1&0&0&-\partial_2
\\
0&0&-1&0&0&0&\partial_1
\\
0&1&0&-1&&-\partial_2&\partial_2
\\
-1&0&1&0&0&\partial_1&-\partial_1
\\
0&0&0&0&0&-m&0
\\
0&0&\partial_2&-\partial_1&m&0&0
\\
\partial_2&-\partial_1&-\partial_2&\partial_1&0&0&0
\end{array}\right)\delta^2(x-y).
\eeq
$F^{(1)}$ exhibits one zero mode
\beq
u^{(1)T}_x(z) = (0,0,\partial_1,\partial_2,0,0,1)\delta^2(x-z).
\label{gaugezeromode1}
\eeq
Noting that
\beq\label{K1vector}
(K^{(1)}_\alpha)= \frac{\delta V^{(1)}}{\delta \xi_\alpha}
= \left(
\begin{array}{c}
 m f^1
\\
 m f^2
\\
-\partial_2f^0
\\
\partial_1f^0
\\
\epsilon_{ij}\partial^iA^j - m f^0
\\
0
\\
0
\end{array}\right),
\eeq we find that the new constraint vanishes identically:
\beq \int d^2z~ u^{(1)}_{\alpha_1,x}(z)\frac{\delta
V^{(1)}}{\delta\xi^{(1)}(z)}= (\partial_1\partial_2 f^0 -
\partial_1\partial_2 f^0) \equiv 0\,.
\eeq
Hence the algorithm ends
at this point. We now write $F^{(1)}$ in the form
\beq
F^{(1)}(x,y)=\left(
\begin{array}{cc}
f&M
\\
-M^T&0
\end{array}\right)\delta^2(x-y),
\eeq
where
\beq\label{fmatrix}
f(x,y)=\left(
\begin{array}{cccccc}
0&0&0&1&0&0
\\
0&0&-1&0&0&0
\\
0&1&0&-1&0&-\partial_2
\\
-1&0&1&0&0&\partial_1
\\
0&0&0&0&0&-m
\\
0&0&\partial_2&-\partial_1&m&0
\end{array}\right)\delta^2(x-y),
\eeq
and $M$ the $1\times 6$ matrix
\beq
M_{\alpha_1}(x,y) = \left(-\frac{\partial\varphi_2(y)}{\partial\xi^{(1)}_{\alpha_1}(x)}\right)=
\left(
\begin{array}{cc}
-\partial_2
\\
\partial_1
\\
\partial_2
\\
-\partial_1
\\
0
\\
0
\end{array}\right)\delta^2(x-y).
\eeq
We next observe that $\det f\not = 0$, so that the inverse of $f$ above exists.
It is readily computed to be
 \beq
 f^{-1}(x,y)=\left(
 \begin{array}{cccccc}
 0&-1&0&-1&{1\over m}\partial_1&0
 \\
 1&0&1&0&{1\over m}\partial_2&0
 \\
 0&-1&0&0&0&0
 \\
 1&0&0&0&0&0
 \\
 -{1\over m}\partial_1&-{1\over m}\partial_2&0&0&0&{1\over m}
 \\
 0&0&0&0&-{1\over m}&0
\end{array}\right)\delta^2(x-y).
\eeq The zero mode (\ref{gaugezeromode1}) is of the general form \cite{wozneto} \beq \label{generalgaugezeromode}
u_{\alpha_1,x}^{(1)}(z)=\left(\sum_{B=1}^6\int d^2y f^{-1}_{AB}(x,y)\frac{\partial
\varphi_2(z)}{\partial\xi^{(1)}_{B}(y)},1\right), \quad A,B = 1,\cdots,6, \eeq where we label the subspace on
which $f^{-1}$ is defined by the indices $A,B,C,\cdots$. As we shall see, in the algorithm of Dirac these label
the complete set of second-class constraints. The zero mode (\ref{gaugezeromode1}) is the generator of gauge
transformations in the sense \cite{wozneto} \beq\label{gaugesymmetry1} \delta\xi^{(1)}_{\alpha_1}(x) = \int d^2z~
u^{(1)}_{\alpha_1,x}(z)\epsilon(z). \eeq

With the aid of (\ref{generalgaugezeromode}) we can readily rewrite this in terms of symplectic brackets. Let $F$
and $G$ be functions of the {\it dynamical} field variables $\xi_A$. We define generalized symplectic structures
by
\beq
\label{starbrackets} \{F,G\}^* = \int d^2z\int d^2z' \frac{\partial F}{\partial \xi_A(z)}
f^{-1}_{AB}(z,z') \frac{\delta G}{\delta \xi_B(z')}. \label{fxgy}
\eeq
In particular
\beq \{\xi_A(x),\xi_B(y)\}^* = f_{AB}^{-1}(x,y),
\eeq
or explicitly
\bea
\{f^i(x),f^j(y)\}^* &=& -\epsilon^{ij}\delta^2(x-y),\nonumber\\
\{f^i(x),A^j(y)\}^* &=& -\epsilon^{ij}\delta^2(x-y),\nonumber \eea in agreement with the Dirac brackets in
Eq.(\ref{Diracbrackets}).

In terms of the symplectic structure  (\ref{starbrackets})
we may write (\ref{gaugesymmetry1}) in a form which will be convenient for later
comparison:
\bea\label{gaugetransformation-general}
(\delta\xi^{(1)}_{\alpha_1}(x)) &=& \left(\int d^2y \int d^2z~ f^{-1}_{AB}(x,y) \frac{\partial \varphi_2(z)}{\partial\xi^{(1)}_{B}(y)}\epsilon(z),\epsilon(x)\right)\nonumber\\
&=& \left(\int d^2z~ \{\xi^{(1)}_A(x),\varphi_2(z)\}^*\epsilon(z),\epsilon(x)\right).
\eea
Explicitly
\bea \delta
A^1 &=& -\partial^1\epsilon, \,\quad \delta A^2 = - \partial^2\epsilon,
\nonumber\\
\delta f^0 &=& 0,\,\quad \delta f^1 = 0,\,\quad \delta f^2 = 0,\,
\nonumber\\
\delta\eta_1 &=& 0,\,\quad \delta\eta_2 = \epsilon. \label{delfields}
\eea
Recalling the redefinition $\dot\eta_2
- A^0\to \dot\eta_2$, we see that $\delta\eta_2 = \epsilon$ implies $\delta A^0 = -\partial^0\epsilon$, in
agreement with our expectations.

\bigskip\noindent
\begin{center}
{\bf Hamiltonian description}
\end{center}

\bigskip
It is instructive to compare the above results with the Hamiltonian description.  Our starting point is again the
first-order Lagrangian (\ref{Lsymplectic}). The canonical momenta conjugate to $\xi_\alpha$ are given by \beq
{\cal P}_\alpha =a_\alpha,\nonumber \eeq and correspondingly we have six primary constraints, which we write in
the canonical form \beq \label{Hamiltonconstraints} \phi_\alpha \equiv {\cal P}_\alpha - a_\alpha \approx 0\,.
\eeq The corresponding primary Hamiltonian governing the time development of the system \cite{dirac64} is thus
given by \bea \label{primaryHamiltonian} H_p &=& \sum_\alpha\int d^2x ~{\cal P}_\alpha\dot\xi_\alpha(x) - L +
\sum_\alpha\int d^2x~ v_\alpha\phi_\alpha
\nonumber\\
&=& V[\xi] + \sum_\alpha \int d^2x~ v_\alpha\phi_\alpha\,, \eea where $v_\alpha$ are Lagrange multipliers (after
suitable redefinition), and $V[\xi]$ is the potential (\ref{pot}).

With the above ``canonical" form for the {\it primary} constraints we have for the corresponding Poisson brackets
\beq \label{primaryPoissonbrackets} \{\phi_\alpha(x),\phi_\beta(y)\} =\frac{\partial
a_\beta(y)}{\partial\xi_\alpha(x)} -  \frac{\partial a_\alpha(x)}{\partial\xi_\beta(y)}\equiv
F^{(0)}_{\alpha\beta}(x,y). \eeq As we have seen, this matrix is not invertible, and possesses in fact ``two" zero
modes\footnote{In order to simplify the language, we do not say ``two-fold infinity" of zero modes.}. They are
obtained as usual by requiring the persistence in time of the primary constraints, and are found to be just
$\varphi_1(z)$ and $\varphi_2(z)$, defined in Eq. (\ref{phi1phi2}). They represent the first generation of
secondary constraints. There are no further (higher generation)  constraints, and the algorithm ends at this
point. We now collect all the constraints into a single ``vector" \beq \label{collectedconstraints} (\Omega_A) =
(\{\phi_{\alpha}\},\varphi_1, \varphi_2). \eeq We correspondingly write for the second-class constraints \beq
\label{secondclassconstraints} (\Omega^{(2)}_{\bar A})= (\{\phi_{\alpha}\},\varphi_1). \eeq The range of values that $\bar A$
takes is implicit in the notation. It is readily recognized that, because of the ``canonical" form of the primary
constraints, \beq \{\Omega_{A}(x),\Omega_{B}(y)\} =\left(\begin{array}{cc}
f(x,y)&(-\frac{\partial\varphi_2(y)}{\partial\xi_{\bar A}(x)})
\\
(\frac{\partial\varphi_2(y)}{\partial\xi_{\bar B}(x)})&0
\end{array}\right)=F^{(1)}(x,y)
\nonumber
\eeq
with
\[
f_{\bar A\bar B}(x,y)=\{\Omega^{(2)}_{\bar A}(x),\Omega^{(2)}_{\bar B}(y)\},
\]
the elements of the matrix (\ref{fmatrix}). Since the submatrix $f$, being constructed from the second-class
constraints, is invertible, we may define the ``Dirac brackets" in the conventional way, as \bea \{F,G\}_{\cal D}
&=&  \{F,G\} -\sum_{\bar A,\bar B=1}^6 \int d^2z'\int d^2z ~\{F,\Omega^{(2)}_{\bar A}(z)\}f^{-1}_{\bar A\bar B}(z,z')
\{\Omega^{(2)}_{\bar B}(z'),G\}\nonumber\\
&=& \int\int d^2z d^2z'~ \frac{\delta F}{\partial\xi_{\bar A}(z)}f^{-1}_{\bar A\bar B}(z,z') \frac{\delta G}{\partial
\xi_{\bar B}(z')}\,.\nonumber \eea Hence the Dirac brackets coincide with the generalized Poisson brackets
(\ref{starbrackets}). In particular choosing for $F$ the coordinates $\xi^{(1)}_{\alpha_1}$, and for $G$ the
generator of the gauge transformation on Hamiltonian level,
\[
G = \int d^2y~\left[\pi^A_0(y)\epsilon_1(y) + \varphi_2(y)\epsilon_2(y)\right],
\]
we recover the gauge transformation (\ref{delfields})
upon using the Lagrangian restriction on the gauge parameters
commented on before.

\section{ Symplectic embedding of master Lagrangian}
\setcounter{equation}{0}
\renewcommand{\theequation}{\arabic{section}.\arabic{equation}}

It is interesting to examine the relation between the BFT embedding procedure
(improved Dirac approach)
discussed in section 2, and a corresponding embedding in the symplectic formulation.
The procedure of section 2 has led to a St\"uckelberg Lagrangian, where
the field $f^\mu$ has been gauged by the introduction of a St\"uckelberg
scalar $\theta$. This suggests the introduction of an additive ``Wess-Zumino"
term to the Lagrangian density (\ref{Lsymplectic}) of the Lorentz covariant form
\[
{\cal L}_{WZ} = \alpha f^\mu\partial_\mu\theta + \frac{\beta}{2}\partial_\mu\theta\partial^\mu\theta.
\]
Following a standard recipe \cite{deser,HKPR} for constructing the corresponding first order Lagrangian, one
readily checks that the corresponding equivalent symplectic Lagrangian now reads as in (\ref{Lsymplectic}), with
\beq \label{xi0-theta} (\xi_\alpha) = (f^1,f^2,A^1,A^2,\theta,\pi_\theta,f^0,A^0), \eeq \beq (a_\alpha) =
\left(-\frac{1}{2}A^2,\frac{1}{2}A^1,-\frac{1}{2}(f^2 - A^2), \frac{1}{2}(f^1 - A^1),\pi_\theta,0,0,0\right), \eeq
and
\bea \label{thetapotential} V[A,f,\theta,\pi_\theta]&=&\int d^2x~\left[-{m\over 2}f_\mu f^\mu
+f^0\epsilon_{ij}\partial^i A^j + A^0(\epsilon_{ij}\partial^i
f^j - \epsilon_{ij}\partial^i A^j )\right.\nonumber\\
&+& \left.\frac{1}{2\beta}(\pi_\theta - \alpha f^0)^2 - \alpha f^i\partial_i\theta -
\frac{\beta}{2}\partial^i\theta\partial_i\theta\right] \,. \eea The Euler-Lagrange equations then read as in
(\ref{EulerLagrange}), with $K^{(0)}$ replaced by \beq \label{Kvector1} (K^{(0)}_\alpha)= \frac{\delta  V}{\delta
\xi_\alpha(x)} = \left(
\begin{array}{c}
\partial^2 A^0 + m f^1 + \alpha\partial^1\theta
\\
-\partial^1 A^0 + m f^2 + \alpha\partial^2\theta
\\
-\partial_2(f^0 - A^0)
\\
\partial_1(f^0 - A^0)
\\
\alpha\partial_if^i + \beta\partial_i\partial^i\theta
\\
\frac{1}{\beta}(\pi - \alpha f^0)
\\
\epsilon_{ij}\partial^iA^j -\frac{\alpha}{\beta}(\pi - \alpha f^0)- m f^0
\\
\epsilon_{ij}\partial^i(f^j - A^j)
\end{array}\right),
\eeq
and $F^{(0)}_{\alpha\beta}$ the
(pre)symplectic form now replaced by
\beq
F^{(0)}(x,y)=\left(
\begin{array}{cccc}
0&\epsilon&0&0
\\
\epsilon&-\epsilon&0&0
\\
0&0&-\epsilon&0
\\
0&0&0&0
\end{array}\right)\delta^2(x-y).
\eeq
As is again evident, since $\det F^{(0)}=0$, the matrix $F^{(0)}$ is not invertible. In fact, the rank of
this matrix is $6$, so that there exist two-fold infinity of zero-generation (left) zero modes
$u^{(0)}(\sigma,z)$, labelled by discrete indices $\sigma=1,2$ and the continuum label $z$, with components:
\bea
u^{(0)T}_{x}(1;z)&=&(0,0,0,0,0,0,-1,0)\delta^2(x-z),
\nonumber\\
u^{(0)T}_{x}(2;z)&=&(0,0,0,0,0,0,0,-1)\delta^2(x-z),\, \eea implying the constraints \bea
\varphi_1(z)&=&-\frac{\delta V}{\delta f^0(z)} =m f^0(z) - \epsilon_{ij}\partial^iA^j(z) +
\frac{\alpha}{\beta}(\pi_\theta(z) - \alpha f^0(z)),
\nonumber\\
\varphi_2(z)&=&-\frac{\delta V}{\delta A^0(z)} =-\epsilon_{ij}\partial^i(f^j(z) - A^j(z)).
\label{gaugedconstraints} \eea We proceed as in section 3, being led to a first-level Lagrangian \beq L^{(1)} =
\int d^2x \sum_{\alpha_1=1}^9 a^{(1)}_{\alpha_1}\dot\xi^{(1)}_{\alpha_1} - V^{(1)}[\xi] \label{firstlag2} \eeq
with a new set of ``first-generation" dynamical variables in which $A^0$ has again been absorbed into a
redefinition of $\eta_2$, \beq (\xi^{(1)}_{\alpha_1}) = (f^1,f^2,A^1,A^2,\theta,\pi_\theta,f^0,\eta_1,\eta_2),
\eeq and \beq \left( a^{(1)}_{\alpha_1}(x) \right) = \left(-\frac{1}{2}A^2,\frac{1}{2}A^1,-\frac{1}{2}(f^2 - A^2),
\frac{1}{2}(f^1 - A^1),\pi_\theta,0,0,-\varphi_1(x),-\varphi_2(x)\right), \eeq and \bea \label{thetapotential2}
V^{(1)}[A,f,\theta,\pi_\theta]&=&\int d^2x~\left[-{m\over 2}f_\mu f^\mu
+f^0\epsilon_{ij}\partial^i A^j\right. \nonumber\\
&+&\left.\frac{1}{2\beta}(\pi_\theta - \alpha f^0)^2 - \alpha f^i\partial_i\theta -
\frac{\beta}{2}\partial^i\theta\partial_i\theta\right].  \eea The equations of motion now take the form
(\ref{firstlevel-eq}), where the ``first-generation" symplectic form $F^{(1)}_{\alpha_1,\beta_1}$ is now given by
(\ref{F1-symplectic}), with \beq \label{F1gauge-symplectic} F^{(1)}(x,y)=\left(
\begin{array}{ccccccccc}
0&0&0&1&0&0&0&0&-\partial_2
\\
0&0&-1&0&0&0&0&0&\partial_1
\\
0&1&0&-1&0&0&0&-\partial_2&\partial_2
\\
-1&0&1&0&0&0&0&\partial_1&-\partial_1
\\
0&0&0&0&0&-1&0&0&0
\\
0&0&0&0&1&0&0&-\frac{\alpha}{\beta}&0
\\
0&0&0&0&0&0&0&-\kappa&0
\\
0&0&\partial_2&-\partial_1&0&\frac{\alpha}{\beta} &\kappa&0&0
\\
\partial_2&-\partial_1&-\partial_2&\partial_1&0&0&0&0&0
\end{array}\right)\delta^2(x-y),\nonumber
\eeq
where $\kappa=m-\alpha^2/\beta$, and $K^{(1)}$ is given by
\beq \label{K1gaugevector} (K^{(1)}_\alpha)=
\frac{\delta V^{(1)}}{\delta \xi_\alpha} = \left(
\begin{array}{c}
m f^1 + \alpha\partial^1\theta
\\
m f^2 + \alpha\partial^2\theta
\\
-\partial_2 f^0
\\
\partial_1 f^0
\\
\alpha\partial_if^i + \beta\partial_i\partial^i\theta
\\
\frac{1}{\beta}(\pi_\theta - \alpha f^0)
\\
\epsilon_{ij}\partial^iA^j -\frac{\alpha}{\beta}(\pi_\theta - \alpha f^0)- m f^0
\\
0
\\
0
\end{array}\right).
\eeq $F^{(1)}$ exhibits one zero mode \beq \label{gaugezeromode2} u^{(1)T}_x(1;z) =
(0,0,\partial_1,\partial_2,0,0,0,0,1)\delta^2(x-z), \eeq implying however an identically vanishing constraint. For
$\frac{\alpha^2}{\beta}= m$ there are two further zero modes: \bea
u^{(1)T}_x(2;z) &=& (0,0,0,0,0,0,1,0,0)\delta^2(x-z),\\
u^{(1)T}_x(3;z) &=& (\partial_1,\partial_2,0,0,\frac{\alpha}{\beta},0,0,1,0)\delta^2(x-z). \label{zeromodes} \eea
The zero modes $u^{(1)T}_x(2;z)$ and $u^{(2)T}_x(3;z)$ both imply the constraint $\varphi_1 = 0$ in
(\ref{gaugedconstraints}), now evaluated for $\kappa=m-\frac{\alpha^2}{\beta}= 0$: \beq \varphi_1 \equiv
\epsilon_{ij}\partial^iA^j - \frac{\alpha}{\beta}\pi_\theta. \eeq The corresponding potential
(\ref{thetapotential2}) now takes the form \beq \label{thetapotential1} V^{(1)}[A,f,\theta,\pi_\theta]=\int d^2x~
\left[-{m\over 2}f_i f^i + \frac{1}{2\beta}\pi_\theta^2 - \alpha f^i\partial_i\theta -
\frac{\beta}{2}\partial^i\theta\partial_i\theta -f^0 \varphi_1\right]. \, \eeq We may thus absorb the term $f^0
\varphi_1$ in the potential into a redefinition of the variable $\eta_1$: $\dot\eta_1- f^0\to \dot\eta_1$.
Correspondingly \beq (\xi^{(1)}_{\alpha_1}) = (f^1,f^2,A^1,A^2,\theta,\pi_\theta,\eta_1,\eta_2), \eeq and
$K^{(1)}$ is given by \beq \label{K1gaugevector1} (K^{(1)}_{\alpha_1})= \frac{\delta V^{(1)}}{\delta
\xi_{\alpha_1}} = \left(
\begin{array}{c}
m f^1 + \alpha\partial^1\theta
\\
m f^2 + \alpha\partial^2\theta
\\
-\partial_2 f^0
\\
\partial_1 f^0
\\
\alpha\partial_if^i + \beta\partial_i\partial^i\theta
\\
\frac{1}{\beta}(\pi_\theta - \alpha f^0)
\\
0
\\
0
\end{array}\right),
\eeq
\beq
\label{F1final}
F^{(1)}(x,y)=\left(
\begin{array}{cc}
f&M
\\
-M^T&0
\end{array}\right)\delta^2(x-y),
\eeq
where
\beq
\label{fsymplectic}
f(x,y)=\left(
\begin{array}{ccc}
0&\epsilon&0
\\
\epsilon&-\epsilon&0
\\
0&0&-\epsilon \nonumber
\end{array}\right),
\eeq
and $M$ is the $2\times 6$ matrix
\beq
M_{\alpha_1}(x,y) =
\left(
\begin{array}{cc}
0&-\partial_2
\\
0&\partial_1
\\
-\partial_2&\partial_2
\\
\partial_1&-\partial_1
\\
0&0
\\
-\frac{\alpha}{\beta}&0
\end{array}\right)\delta^2(x-y).\nonumber
\eeq
In the form (\ref{F1final}), $F^{(1)}$ still has two zero modes:
\bea
\label{gauge-zeromodes}
u^f_{x}(z) &=& (\partial_1,\partial_2,0,0,\frac{\alpha}{\beta},
0,1,0)\delta^2(x-z),\nonumber\\
u^A_{x}(z) &=& (0,0,\partial_1,\partial_2,0,0,0,1)\delta^2(x-z). \eea One readily checks that they imply
identically vanishing constraints:
\[
u^f(z)\cdot K^{(1)} \equiv 0\,,
\quad u^A(z)\cdot K^{(1)} \equiv 0\,.
\]
They therefore generate the following gauge transformation:
\beq
\label{gaugetransformation}
\delta\xi_{\alpha_1}(x) = \int d^2z~\left[u^f_{\alpha_1,x}(z)\epsilon^f(z) +
u^A_{\alpha_1,x}(z)\epsilon^A(z)\right], \eeq or explicitly
\bea \delta A^1 &=& -\partial^1\epsilon^A, \, \quad
\delta A^2 = -\partial^2\epsilon^A,
\nonumber\\
\delta f^1 &=&  -\partial^1\epsilon^f, \,\quad
\delta f^2 = -\partial^2\epsilon^f,\nonumber\\
\delta \theta &=& \frac{\alpha}{\beta}\epsilon^f,\,\quad \delta \pi_\theta = 0,\nonumber\\
\delta\eta_1 &=& \epsilon^f,\,\quad \delta\eta_2 = \epsilon^A. \eea Recalling the relabelling $\dot\eta_1 - f^0\to
\dot\eta_1$ and $\dot\eta_2 - A^0\to\dot\eta_2$, we see that the transformations for $\eta_i$ imply
\[
\delta f^0 = -\partial^0\epsilon^f,\,\quad \delta A^0 = -\partial^0\epsilon^A
\]
in accordance with our expectations: $\delta f^\mu= -\partial^\mu \epsilon^f$, $\delta A^\mu= -\partial^\mu
\epsilon^A$, and $\delta\theta=\epsilon^f$ which are exactly the same as the transformation
(\ref{idqm-transformation}) obtained from the improved DQM when we assign the coefficients for $\alpha=\beta=m$.

\bigskip\noindent
\begin{center}
{\bf Hamiltonian point of view}
\end{center}

\bigskip
>From the Hamiltonian point of view, the symplectic Lagrangian (\ref{Lsymplectic}) implies again the primary
constraints (\ref{Hamiltonconstraints}) with (\ref{xi0-theta}) -- (\ref{thetapotential2}), as well as two
additional primary constraints
\[
\phi^\theta \equiv {\cal P}_\theta - \pi_\theta \approx 0,
\]
and
\[
\phi^{\pi_\theta} \equiv {\cal P}_{\pi_\theta} \approx 0,
\]
where ${\cal P}^\theta, {\cal P}^{\pi_\theta}$ are the momenta conjugate to $\theta$ and $\pi_\theta$,
respectively. The canonical Hamiltonian has the characteristic feature of being
just given by the symplectic potential. Hence we have for the
primary Hamiltonian
\[
H_p = V[A,f,\pi_\theta,\theta] + \int d^2z \sum_\sigma v_\alpha \phi_\alpha.
\]
The constraints $\phi_i^A \approx 0, \phi_i^f \approx 0, \phi^\theta \approx 0, \phi^\pi \approx 0$ fix the
Lagrange multipliers $v^i_A = 0, v^i_f = 0, v_\theta = 0, v_\pi = 0$:
\bea v_i^f &=& -m \epsilon_{ij}f^j - \partial_{i}f^0 - \alpha\epsilon_{ij}\partial^j\theta,\nonumber\\
v_i^A &=& -m\epsilon_{ij}f^j-\partial_i A^0 - \alpha\epsilon_{ij}\partial^j\theta
,\nonumber\\
v^\theta &=& \frac{1}{\beta}(\pi - \alpha f^0),\nonumber\\
v^\pi &=& -\alpha\partial_if^i - \beta\partial_i\partial^i\theta.
\eea

We recognize that the elements of $F^{(0)}$ are just the Poisson brackets of the primary constraints:
$F^{(0)}_{\alpha\beta} = \{\phi_\alpha,\phi_\beta\}$. The usual Dirac algorithm leads to the secondary constraints
$\varphi_a \approx 0, a=1,2$, which for $\alpha^2/\beta= m$ are both found to have identically vanishing Poisson
brackets with the primary Hamiltonian, after making use of the explicit expressions for the fixed Lagrange
parameters. Hence no new constraints are generated, and we have in the final stage two first-class primary, and
two first-class secondary constraints, generating in the usual manner the extended gauge symmetry of the
Hamiltonian. It is interesting that in the symplectic approach we directly obtain the more restricted symmetry of
the Lagrangian.

\section{Conclusion}

It has been the primary objective of this paper to illustrate in terms of a non-trivial model as described by the
master Lagrangian of Deser and Jackiw \cite{deser}, how the embedding of Hamiltonian systems with first and
second-class constraints into an extended gauge theory is realized in the context of the ``improved" Dirac
quantization (BFT) approach on the one hand, and the ``improved" symplectic approach, on the other.  Rather than
proceeding iteratively as one does in the improved DQM approach, we have simplified the calculation in the
symplectic case by making use of manifest Lorentz invariance in our {\it ansatz} for the Wess-Zumino (WZ) term to
be added to the master Lagrangian, and then reformulating the problem in terms of an equivalent first order
Lagrangian. We have further established a one-to-one correspondence between the symplectic and the Dirac approach.
Just as in the case of the improved DQM procedure, the symplectic embedding procedure requires the introduction of
an even number of additional fields, which, following the Faddeev-Jackiw prescription~\cite{jackiw85} can be
chosen to be canonically conjugate pairs. This is in line with the fact that the number of second-class
constraints is always even, and that the BFT embedding procedure requires that phase space be augmented by one
degree of freedom for each secondary constraint. This fact has not been recognized in a recent paper on the
subject~\cite{neto0109}, where in our notation, $\pi_{\theta}$ has effectively been taken to be a function of
$A^{i}$, $\pi_{i}$ and  $\theta$.


\vskip 1.0cm
One of us (KDR) would like to thank the Sogang High Energy Physics
Group for their warm hospitality. Two of us (STH and YJP)
acknowledge financial support from the Korea Research Foundation,
Grant No. KRF-2001-DP0083. \\

\begin{appendix}

\begin{center}
{\bf Appendix: Improved Dirac Quantization Method}
\end{center}

\setcounter{equation}{0}
\renewcommand{\theequation}{A.\arabic{equation}}


In this appendix we demonstrate how the improved DQM can be used in order to turn the model defined by
the Master Lagrangian into a fully first-class system on the Hamiltonian level. In order to extract the true
second-class constraints, we redefine the secondary constraint $\varphi^A$ as follows: \beq
\omega^{A}=\pa^{i}\pi_{i}^{A}-\frac{1}{2}\epsilon_{ij}\pa^{i}f^{j} +\frac{1}{2}\epsilon_{ij}\pa^{i}A^{j}.
\label{const33} \eeq The nonvanishing Poisson brackets are then given as \bea \{\phi_0^{f}(x),
\varphi^{f}(y)\}&=&-m\delta^2(x-y),
\nonumber\\
\{\phi_{i}^{f}(x), \phi_{j}^{A}(y)\}&=&\epsilon_{ij}\delta^2(x-y),
\nonumber\\
\{\phi_{i}^{A}(x), \phi_{j}^{A}(y)\}&=&-\epsilon_{ij}\delta^2(x-y), \label{brackets1} \eea which show that
$\phi_{0}^{A}$ and $\omega^{A}$ are first-class.

Redefining the constraints:
\begin{equation}
\label{redef-con} (\Omega_{1},\Omega_{2},\Omega_{3},\Omega_{4},\Omega_{5},
\Omega_{6})=(\phi_{0}^{f},\varphi^{f},\phi_{1}^{f},\phi_{2}^{f},\phi_{1}^{A}, \phi_{2}^{A}),
\end{equation}
we obtain the second-class algebra \beq \Delta_{\alpha\beta}=\{\Omega_{\alpha},\Omega_{\beta}\}= \left(
\begin{array}{ccc}
-m \epsilon &0 &0 \\
0 &0  & \epsilon\\
0 &  \epsilon &-\epsilon\\
\end{array}
\right)\delta^2(x-y), \label{diracs} \eeq where $\epsilon$ is the Levi-Civita tensor with $\epsilon_{12}=1$ and
$0$ is the $2\times 2$ null matrix.

The consistent quantization of the self-dual model is then
obtained in terms of the following nonvanishing Dirac brackets
\begin{equation}
\label{dbrackets}
\begin{array}{ll}
\{f^0 (x), f^i(y)\}_{D}=-\frac{1}{m}\partial^{x}_{i}\delta^2(x-y),
&\{f^i (x), f^j (y) \}_{D}=-\epsilon^{ij}\delta^2(x-y),\\
\{f^i (x), A^j (y) \}_{D}=-\epsilon^{ij}\delta^2(x-y),
&\{A^i (x), \pi_{j}^{A}(y) \}_{D}=\frac{1}{2}\delta^{i}_{j}\delta^2(x-y),\\
\{\pi_{i}^{A}(x), \pi_{j}^{A} (y) \}_{D}=\frac{1}{4}\epsilon_{ij}\delta^2(x-y),
&\{\pi_{i}^{f}(x), \pi_{j}^{A}(y) \}_{D}=-\frac{1}{4}\epsilon_{ij}\delta^2(x-y),\\
\{f^i (x), \pi_j^{f} (y) \}_{D}=\frac{1}{2}\delta^{i}_{j} \delta^2(x-y),
&\{f^0 (x), \pi_{i}^{A} (y) \}_{D}=-\frac{1}{2m}\epsilon_{ij}\pa^{j}_{x} \delta^2(x-y), \\
\{A^0 (x), \pi_{0}^{A} (y) \}_{D}=\delta^2(x-y).&
\end{array}
\label{diracs2}
\end{equation}


Now, let us extend phase space further to embed all the second-class constraints into the corresponding
first-class ones, while in section 2 partially embed them by eliminating the second-class ones originated from the
symplectic structure of the Chern-Simons term.

To embed all the second-class constraints into the first-class ones by following the improved DQM as in section 2,
we first introduce three pairs of auxiliary fields such as $(\theta^{1},\theta^{2})$, $(\sigma^{1},\sigma^{2})$
and $(\rho^{1},\rho^{2})$ satisfying the canonical Poisson brackets
\begin{equation}
\{\theta^{1}(x),\theta^{2}(y) \} = \{\sigma^{1}(x),\sigma^{2}(y)\}=\{\rho^{1}(x),\rho^{2}(y)\}=\delta^2(x-y),
\label{auxil}
\end{equation}
which define $\omega^{\alpha\beta}$ in (\ref{intro-aux}) as
\begin{eqnarray}
\label{omegam} \omega^{\alpha\beta}= \left(
\begin{array}{ccc}
\epsilon & 0 & 0
\\
0 & \epsilon & 0
\\
0 & 0 & \epsilon
\end{array}\right).
\end{eqnarray}
>From the strong involution relations $\{\tilde{\Omega}_\alpha, \tilde{\Omega}_\beta\}=0$ and the ansatz of the
form (\ref{ansatzlinear}), we obtain a solution $X_{\alpha\beta}$ explicitly as
\begin{eqnarray}
X_{\alpha\beta}= \left(
\begin{array}{cccccc}
\sqrt{m} & 0 & 0 & 0 & 0 & 0
\\
0 & \sqrt{m} & 0 & 0 & 0 & 0
\\
0 & 0 & -1 & 0 & -1 & 0 \\
0 & 0 & 0 & 1 & 0 & -1 \\
0 & 0 & 0 & 0 & 1 & 0 \\
0 & 0 & 0 & 0 & 0 & 1
\end{array}\right).
\end{eqnarray}
As results, we converts all the second-class constraints to the first-class ones \bea
\tilde{\Omega}_{i}&=&\Omega_{i}+\sqrt{m}\theta^{i},
\nonumber\\
\tilde{\Omega}_{i+2}&=&\Omega_{i+2}+(-1)^i\sigma^{i} -\rho^{i},
\nonumber\\
\tilde{\Omega}_{i+4}&=&\Omega_{i+4}+\rho^{i}, \label{consttilast} \eea with $i,j=1,2$ satisfying the rank-zero
algebra: $\{\tilde{\Omega}_{\alpha}, \tilde{\Omega}_{\beta} \}=0$.

Similarly, we obtain for the improved first-class fields in the extended phase space \beq
\begin{array}{ll}
\tilde{f}^{0}=f^{0}+\frac{1}{\sqrt{m}}\theta^{2},~~~ &\tilde{f}^{i}=f^{i}+\frac{1}{\sqrt{m}}\pa^{i}\theta^{1}
+(-1)^i\epsilon_{ij}\sigma^{j},\\
\tilde{A}^{0}=A^{0},~~~ &\tilde{A}^{i}=A^{i}+(-1)^i\epsilon_{ij}\sigma^j+\epsilon_{ij}\rho^{j},\\
\tilde{\pi}_{0}^{f}=\pi_{0}^{f}+\sqrt{m}\theta^{1},~~~ &\tilde{\pi}_{i}^{f}=\pi_{i}^{f}+\frac{1}{2}(-1)^i\sigma^i
-\frac{1}{2}\rho^{i},\\
\tilde{\pi}_{0}^{A}=\pi_{0}^{A},~~~
&\tilde{\pi}_{i}^{A}=\pi_{i}^{A}-\frac{1}{2\sqrt{m}}\epsilon_{ij}\pa^{j}\theta^{1}+\frac{1}{2}\rho^{i}.
\end{array}\label{apitildelast} \eeq
>From these, it can readily be shown that in the master self-dual model the Poisson brackets in the extended phase
space are exactly equivalent to the Dirac brackets (\ref{dbrackets})~\cite{gr,KimParkRothe}.

On the other hand, since an arbitrary functional of the improved first-class fields is also first-class, we can
also directly obtain the desired first-class Hamiltonian\footnote{Similar to the section 2, we can also construct
the involutive Hamiltonian $\tilde{H}$ directly by following the method of Batalin et al.~\cite{batalin87} and
thus demonstrate the equivalence of $\tilde{\cal H}_p$ up to total derivatives.} $\tilde{\cal H}$ corresponding to
the Hamiltonian ${\cal H}_p$ in Eq. (\ref{hprimary}) via the substitution $f^\mu\rightarrow\tilde{f}^\mu$,
$A^\mu\rightarrow\tilde{A}^\mu$, $\pi_\mu^{f}\rightarrow\tilde{\pi}_\mu^{f}$ and
$\pi_\mu^{A}\rightarrow\tilde{\pi}_\mu^{A}$: \bea \tilde{\cal H}_{p}&=&{\cal H}_c
+\sqrt{m}\theta^1\pa_if^{i}+\theta^2(-\sqrt{m}f^{0}+\frac{1}{\sqrt{m}}\epsilon_{ij}\pa^iA^j)
+mf^i(-1)^i\epsilon_{ij}\sigma^j
\nonumber\\
& &-(-1)^if^0\pa_i\sigma^i+(f^0-A^0)\pa_i\rho^i -\sqrt{m}\theta^1(-1)^i\epsilon_{ij}\pa^i\sigma^j
 \nonumber\\
&&+\theta^{2}\left(-\frac{1}{2}\theta^2
+\frac{1}{\sqrt{m}}\pa_{i}\rho^{i} -(-1)^i\frac{1}{\sqrt{m}}\pa_{i}\sigma^{i}
\right)+\frac{m}{2}(\sigma^i)^2-\frac{1}{2}\pa_i\theta^1\pa^i\theta^1.\nonumber\\
 \label{htotalt}
\eea

Next, we construct the Hamilton equations of motion for these first-class fields \beq
\begin{array}{ll}
\frac{{\rm d}}{{\rm d}t}\ti{f}^{0}=\pa^{i}\ti{f}^{i},
&\frac{{\rm d}}{{\rm d}t}\ti{f}^{i}=-m\epsilon^{ij}\ti{f}^{j}+\frac{1}{m}\pa^{i}(\epsilon_{jk}\pa^{j}\ti{A}^{k}),\\
\frac{{\rm d}}{{\rm d}t}\ti{A}^{0}=0, &\frac{{\rm d}}{{\rm d}t}\ti{A}^{i} =-
m\epsilon^{ij}\ti{f}^{j}+\pa^{i}\ti{A}^{0}.
\end{array}
\label{eoms} \eeq Here note that the equation of motion for $\ti{f}^{0}$ in Eq. (\ref{eoms}) can be rewritten in
terms of covariant form, $\pa_{\mu}\ti{f}^{\mu}=0$, from which one can obtain the explicit form for
$\ti{f}^{\mu}$, the duality relation~\cite{deser}:
$\ti{f}^{\mu}=\frac{1}{m}\epsilon^{\mu\nu\rho}\pa_{\nu}\ti{A}_{\rho}$, now written in terms of the improved first-class
(gauge invariant) fields in the extended phase space.

\end{appendix}



\begin{thebibliography}{99}
\bibitem{dirac64} P. A. M. Dirac, {\it Lectures on quantum mechanics}
     (Belfer graduate School, Yeshiba University Press, New York, 1964).
\bibitem{becci76} C. Becci, A. Rouet and R. Stora, Ann. Phys. (N.Y.)
           {\bf 98}, 287 (1976);
           I. V. Tyutin, Lebedev Preprint 39 (1975).
\bibitem{kugo79} T. Kugo and I. Ojima,
           Prog. Theor. Phys. Suppl. {\bf 66}, 1 (1979).
\bibitem{fradkin75} E. S. Fradkin and G. A. Vilkovisky,
            Phys. Lett. {\bf B55}, 224 (1975).
\bibitem{henneaux85} M. Henneaux, Phys. Rept. {\bf 126}, 1 (1985).
\bibitem{batalin87} I. A. Batalin and E. S. Fradkin,
    Nucl. Phys. {\bf B279}, 514 (1987); Phys. Lett. {\bf B180}, 157 (1986);
    I. A. Batalin and I. V. Tyutin, Int. J. Mod. Phys. {\bf A6}, 3255 (1991).
\bibitem{idqm} R. Banerjee, Phys. Rev. {\bf D48}, R5467 (1993);
               W. T. Kim and Y.-J. Park, Phys. Lett. {\bf B336}, 376 (1994);
               S.-T. Hong, W. T. Kim and Y.-J. Park,
              Mod. Phys. Lett. {\bf A15}, 1915 (2000); S.-T. Hong, Y.-J. Park,
              K. Kubodera and F. Myhrer, Mod. Phys. Lett. {\bf A16}, 1361
              (2001).
\bibitem{KimRothe} Y.-W. Kim and K. D. Rothe, Nucl. Phys. {\bf B511},
                   510 (1998).
\bibitem{nonPro} R. Banerjee and J. Barcelos-Neto, Nucl. Phys. {\bf B499}, 453
              (1997); M.-I. Park and Y.-J. Park, Int. J. Mod. Phys. {\bf A13},
              2179 (1998).
\bibitem{gafn}  S. Ghosh, Phys. Rev. {\bf D49}, 2990 (1994);
J. G. Zhou, Y. G. Miao, and Y. Y. Liu, Mod. Phys. Lett. {\bf A9}, 1273 (1994); R. Amorim and J. Barcelos-Neto,
Phys. Rev. {\bf D53}, 7129 (1996); M. Fleck and H. O. Girotti, Int. J. Mod. Phys. {\bf A14}, 4287 (1999); C. P.
Nativiade and H. Boschi-Filho, Phys. Rev. {\bf D62}, 025016 (2000).
\bibitem{fujiwara90} T. Fujiwara, Y. Igarashi and J. Kubo,
            Nucl. Phys. {\bf B341}, 695 (1990);
            Phys. Lett. {\bf B251}, 427 (1990); J. Feinberg and M. Moshe,
            Ann. Phys. {\bf 206}, 272 (1991);
            S. Hamamoto, Prog. Theor. Phys. {\bf 96}, 639 (1996).
\bibitem{kim92} Y.-W. Kim, S.-K. Kim, W. T. Kim, Y.-J. Park,
           K. Y. Kim and Y. Kim, Phys. Rev. {\bf D46}, 4574 (1992).
\bibitem{banerjee94} R. Banerjee, H. J. Rothe and K. D. Rothe,
           Phys. Rev. {\bf D49}, 5438 (1994); Nucl. Phys. {\bf B426}, 129
           (1994); Phys. Rev. {\bf D55}, 6339 (1999).
\bibitem{kim95} Y.-W. Kim, Y.-J. Park, K. Y. Kim and Y. Kim, Phys. Rev.
{\bf D51}, 2943 (1995).
\bibitem{KimParkRothe} Y.-W. Kim, Y.-J. Park and K. D. Rothe,
                  J. Phys. {\bf G24}, 953 (1998).
\bibitem{banerjee95ap} N. Banerjee, R. Banerjee and S. Ghosh, Ann. Phys.
            {\bf 241}, 237 (1995).
\bibitem{kkky} Y.-W. Kim, M.-I. Park, Y.-J. Park and S.-J. Yoon,
               Int. J. Mod. Phys. {\bf A12}, 4217 (1997).
\bibitem{phyrep} S.-T. Hong and Y.-J. Park, Phys. Rept. {\bf 358}, 143 (2002),
and references therein.
\bibitem{faddeev86} L. D. Faddeev and S. L. Shatashivili,
                    Phys. Lett. {\bf B167}, 225 (1986);
               O. Babelon, F. A. Shaposnik and C. M. Vialett,
                  Phys. Lett. {\bf B177}, 385 (1986);
             K. Harada and I. Tsutsui, Phys. Lett. {\bf B183}, 311 (1987).
\bibitem{wess71} J. Wess and B. Zumino, Phys. Lett. {\bf B37}, 95 (1971).
\bibitem{jackiw85} L. Faddeev and R. Jackiw, Phys. Rev. Lett. {\bf 60}, 1692 (1988);
    R. Jackiw, {\it Diverse Topics in Theoretical and Mathematical Physics},
     (World Scientific, Singapore, 1995).
\bibitem{wozneto} J. Barcelos-Neto and C. Wotzasek, Mod. Phys. Lett. {\bf A7},
            1172 (1992); Int. J. Mod. Phys. {\bf A7}, 4981 (1992);
            H. Montani and C. Wotzasek, Mod. Phys. Lett. {\bf A8},
           3387 (1993); H. Montani, Int. J. Mod. Phys. {\bf A8}, 3419 (1993).
\bibitem{kimjkps1} Y.-W. Kim, Y.-J. Park, K. Y. Kim, Y. Kim and C.-H. Kim,
            J. Korean Phys. Soc. {\bf 26}, 243 (1993);
            Y.-W. Kim, Y.-J. Park, K. Y. Kim and Y. Kim,
            J. Korean Phys. Soc. {\bf 27}, 610 (1994).
\bibitem{kimjkps2} S.-K. Kim, Y.-W. Kim, Y.-J. Park, Y. Kim, C.-H. Kim and
             W. T. Kim, J. Korean Phys. Soc. {\bf 28}, 128 (1995).
\bibitem{neto0109} A. C. R. Mendes, J. Ananias Neto, W. Oliveira, C. Neves
                   and D. C. Rodrigues, hep-th/0109089.
\bibitem{HKPR} S.-T. Hong, Y.-W. Kim, Y.-J. Park and K. D. Rothe,
hep-th/01112170.
\bibitem{deser} S. Deser and R. Jackiw, Phys. Lett. {\bf B139}, 371 (1984).
\bibitem{BanerjeeRothe} R. Banerjee, H. J. Rothe and K. D. Rothe,
                        Phys. Rev, {\bf D52}, 3750 (1995).
\bibitem{gr} H. O. Girotti and K. D. Rothe, Int. J. Mod. Phys. {\bf A4},
             3041 (1989).
\bibitem{faddev}  L. D. Faddeev, Theor. Math. Phys. {\bf 1}, 1 (1970); P.
Senjanovich, Ann. Phys. (N.Y.) {\bf 100}, 277 (1976).
\bibitem{brr} R. Banerjee, H. J. Rothe and K. D. Rothe, Phys. Lett. {\bf B463},
              248 (1999); {\it ibid.}, {\bf B479}, 429 (2000).
\end{thebibliography}
\end{document}